\documentclass[preprint,8pt]{elsarticle}

\usepackage{amssymb}
\usepackage{graphicx}
\usepackage{amssymb}
\usepackage{graphicx}
\usepackage{amsfonts}
\usepackage{geometry}
\usepackage{epsfig}
\usepackage{CJK}
\usepackage{amsmath}
\usepackage{epstopdf}
\usepackage{mathrsfs}
\usepackage{bbding}
\usepackage{cases}
\usepackage{subfigure}
\usepackage{float}
\usepackage{lineno,hyperref}
\usepackage{lineno,hyperref}
\usepackage{lipsum}
\usepackage{amssymb}
\usepackage{graphicx}
\usepackage{amsfonts}
\usepackage{geometry}
\usepackage{epsfig}
\usepackage{CJK}
\usepackage{amsmath}
\usepackage{graphicx}
\usepackage{epstopdf}
\usepackage{mathrsfs}
\usepackage{bbding}
\usepackage{cases}
\usepackage{subfigure}
\usepackage{float}
\usepackage{lineno,hyperref}
\modulolinenumbers[5]

\journal{}

\begin{document}

\begin{frontmatter}



\title{General high-order rogue wave to NLS-Boussinesq equation with the dynamical analysis}
\author[]{Xiaoen Zhang$^{1, 2}$}
\author[]{Yong Chen$^{1, 2}$\corref{mycorrespondingauthor}}
\cortext[mycorrespondingauthor]{Corresponding author.}
\ead{ychen@sei.ecnu.edu.cn}
\address{1 Shanghai Key Laboratory of Trustworthy Computing, East China Normal University, Shanghai, 200062, China\\
2 MOE International Joint Lab of Trustworthy Software, East China Normal University, Shanghai, 200062, China}

\author{}

\address{}
\date{}
\begin{abstract}
General high-order rogue waves of the NLS-Boussinesq equation are obtained by the KP-hierarchy reduction theory. These rogue waves are expressed with the determinants, whose entries are all algebraic forms. It is found that the fundamental rogue waves can be classified three patterns: four-petals state, dark state, bright state by choosing different parameter $\alpha$ values. As the evolution of the parameter $\alpha$. An interesting phenomena is discovered: the rogue wave changes from four-petals state to dark state, whereafter to the bright state, which are consistent with the change of the corresponding critical points to the function of two variables. Furthermore, the dynamics of second-order and third-order rogue waves are presented in detail, which can be regarded as the nonlinear superposition of the fundamental rogue waves.
\end{abstract}

\begin{keyword}
High-order Rogue wave, NLS-Boussinesq equation, KP reduction technique



\end{keyword}

\end{frontmatter}


\section{Introduction}
\label{}
Recently, there are many studies about the rogue waves on both experimental observation and theoretical analysis, which initially turn to describe the spontaneous ocean surface waves. The possible mechanisms to the generation of rogue wave contain modulation instability\cite{Onorato-2006-PRL,Baronio-2014-PRL}, the interaction between the ripples\cite{Peterson-2003-NPG}, the nonlinear focusing of the transient frequency modulation wave\cite{Pelinovsky-2008-EJM} and so on. The study of rogue wave is currently one of the hot topic encompassing many aspects, such as optics\cite{Solli-Nature-2007,Pierangeli-2015-PRL,Akhmediev-2013-JO}, Bose-Einstein condensates\cite{Bludov-2009-PRA}, plasma\cite{Moslem-2011-PP} and even the finance\cite{Yan-2011-PLA}. Fundamentally, rogue wave is modeled as a transient wavepacket localized in both space and time, aside from having a peak amplitude more than twice the background wave, it has a special feature with the instability and unpredictability. As a one-dimensional integrable scalar equation to display the nonlinear wave propagation, with various applications related to the deep water hydrodynamics to nonlinear optics, the nonlinear Schr\"{o}dinger equation(NLS)\cite{Ohta-2012-PRS,Akhmediev-2009-PRE,Ling-2012-STU, Ling-2014-PRE,He-2011-JPA} play a key role in the description of rogue wave. In 1983, Peregrine\cite{Peregrine-1983-JAMS} first gave a rational rogue wave to the NLS equation, its generation principle is the evolution of the breather when the period tends to infinity and it has been carried out in the optics experiment and water wave tank on the experimental. It is well known that rogue wave has different spatial-temporal structures, for example, the eye-shaped pattern can be found in the scalar equation while the vector ones can admit anti-eye-shaped and four-petals\cite{Zhao-2014-PRE}. Moreover, the corresponding high order rogue wave can be regarded as the nonlinear superposition of the fundamental rogue wave and will exhibit higher peak amplitude. By using the Darboux transformation, the significant fourth order rogue wave was provided on the theoretically by Akhmediev\cite{Akhmediev-2009-PRE}. Chabchoub observe the fifth-order rogue wave in the water tank. Therefore, the understanding of the fundamental one-order rogue wave is crucial to the dynamics property.

Another significant research on the rogue wave is the multicomponent coupled system, which can appear some more exciting dynamical characters including three kinds of interaction: the interaction between the same sates, the interaction between the two different states even the interaction between three different states, due to the additional freedom degrees. For instance, for the coupled NLS equation\cite{Guo-2011-CPL, Zhang-2017-PRE}, it appears the dark rogue wave, the interaction between the rogue wave and other localized waves due to the relative velocity cannot be affected by any trivial changes. The interactions between high-order rogue wave and other localized waves are presented to the cubic-quintic NLS equation\cite{Xu-2017-CPB} by using the Darboux transformation. Apart from the multicomponent coupled system, the high-dimensional system can describe the rogue wave more verisimilitude, we found the rogue wave aroused by the lump soliton and a pair of resonance stripe solitons, which is different from the traditional high-dimensional line rogue wave and it refers to the KP equation\cite{Zhang-2017-arXiv}, the Jimbo-Miwa equation\cite{Zhang-2017-CNSNS}, (2+1)-dimensional KdV equation\cite{Zhang-2017-Non}.

The KP reduction technique is very powerful to research the integrable system and derive the localized waves. In 1980's, it was established by the Kyoto School\cite{Jimbo-1983-EMS,Ohta-1989-RSK} to get the soliton solution of integrable system. Based on this technique, in 2012, Ohta et.al. obtained the $N$-dark-dark soliton solution of a two-coupled NLS equation\cite{Ohta-2011-STD}. In 2014, Feng constructed the general bright-dark soliton solutions coupled with all combinations to the NLS equation\cite{Feng-2014-JPA}. Recently, we studied the Yajima-Oikawa system and Mel'nikov system through the KP reduction technique. In terms of the Yajima-Oikawa equation\cite{Chen-2015-JPSJ,Chen-2015-JPSJ1,Chen-2016-JPA,Chen-2017-arXiv,Chen-2015-PLA}, not only the its multi-dark soliton solution, mixed multi-soliton solutions, high-order rogue wave and rational solutions to the continue case, but also the soliton solution to the semi-discretization coupled case are all obtained. Similarly, the $N$-dark soliton solution, bright-dark mixed $N$-soliton solutions to the Mel'nikov systems are given\cite{Han-2017-JPSJ,Han-2017-JPSJ1}.

Based on the KP reduction technique, we construct the general high-order rogue waves to the NLS-Boussinesq equation,
\begin{equation}\label{zxe0801}
\begin{split}
&i\Phi_t-\Phi_{xx}-u\Phi=0,
\\&u_{xx}+u_{xxxx}+3(u^2)_{xx}-3u_{tt}+(\Phi\Phi^*)_{xx}=0,
\end{split}
\end{equation}
where $\Phi$ is a complex function, $\Phi^*$ is the complex conjugate function of $\Phi$,  and $u$ is a real function, this system is used to describe the nonlinear propagation with the coupled Langmuir and dust-acoustic waves, including some ions, electrons and massive charged dust particles. For slow modulations, the amplitude of Langmuir-wave can be dominated by the NLS equation, while for the small but finite amplitude ion-acoustic wave, it must be governed by a driven Boussinesq equation to act a bidirectional propagation. Generally speaking, NLS equation can be described the Langmuir wave and the linear equation can be used to depict the dust-acoustic waves accompanying with small-amplitude. However, when the propagation wave is located in the near of the dust-acoustic speed with a finite amplitude, the linear equation will not describe this phenomena precisely, conversely, this phenomena can be governed by the nonlinear Boussinesq equation. This equation was first be given in Ref \cite{Rao-1989-JPA}, then its analysis solution, $N$-soliton solutions and other patterns were reported in \cite{Singh-1998-JPP,Hase-1988-JPSJ,Mu-2012-JPSJ}.

This paper, we mainly discuss the general high-order rogue wave to the NLS-Boussinesq equation by using the KP reduction technique. To our knowledge, the high-order rogue waves are never obtained. The obtained rogue wave can be classified as three patterns, four-petals state, dark state and bright state. As the evolution of the parameter $\alpha$, the rogue wave turns from the four-petals state to the dark state until to the bright state, the progress can be explained through the critical points property to the function with two-variables, that is, if there are four critical points, two are the maximums and two are the minimums, it is the four-petals state; it there are three critical points, one is the minimums and two are the maximums, it is the dark state, otherwise, one is the maximums and two are minimums, it is the bright state. The rich dynamical behaviors are plotted. In Ref.\cite{Zhao-2014-PRE1}, the author demonstrated that these rogue waves in the two-coupled NLS equation can be transited to each other, in other words, the four-petals rogue wave can be changed into the bright state or the dark state with the varying of the relative frequency. However, in our paper, the four-petals state cannot be changed into the bright, perhaps the reason is that the relative amplitude of these two minimums are bigger than these two maximums no matter how the parameter $\alpha$ transform.

The structure of this paper is organized as follow: section 2 list some KP hierarchy with Gram determinant and reduce them into the bilinear equation of the NLS-Boussinesq equation, in which the elements of this equation are algebraic expression. The dynamical one-order fundamental rogue wave, high-order rogue wave solutions and the corresponding analysis are presented in section 3. The last section is the conclusion and some summary.
\section{The bilinear form of NLS-Boussinesq equation derived from the KP hierarchy}
We want to obtain the bilinear form of Eq.(\ref{zxe0801}) connected with the famous KP hierarchy in this section, which is crucial to derive the fundamental rogue wave and high-order rogue wave. Let us introduce the following variable transformation
\begin{equation*}
\begin{split}
&\Phi=e^{\rm i(\alpha x+\alpha^2t)}\frac{g}{f},
\\&u=2\frac{\partial^2}{\partial x^2}\mbox{log} f
\end{split}
\end{equation*}
where $\alpha$ is an arbitrary real constant, $f$ is a real-value function and $g$ is a complex-valued function. Under this transformation, the NLS-Boussinesq equation can be transferred its bilinear form as
\begin{equation}
\begin{split}
&\left(\mbox{i} D_t-2\mbox{i}\alpha D_x-D_x^2\right)g\cdot f=0,\\
&\left(D_x^2+D_x^4-3D_t^2-1\right)f\cdot f+gg^*=0,
\end{split}
\end{equation}
where $g^*$ is the complex conjugate and $D$ is the  bilinear operator defined as
\begin{equation*}
D_t^mD_x^n(a,b)\equiv\frac{\partial^m}{\partial s^m}\frac{\partial^m}{\partial y^m}a(t+s, x+y)b(t-s,x-y)\big|_{s=0,y=0}.
\end{equation*}
By means of the famous KP theory, it needs to some skills to obtain the polynomial solutions of $f$ and $g$ by using the $\tau$ function of KP hierarchy under the reduction. We will give prominence to the steps of the derivation in detail, which is shown in the next.

Firstly, we list three bilinear equations from the KP hierarchy
\begin{equation}\label{zxe0803}
\begin{split}
&\left(D^{2}_{x_1}+2aD_{x_1}-D_{x_2}\right)\tau(k+1,l)\cdot\tau(k,l)=0,\\
&\left(\frac{1}{2}D_{x_1}D_{x_{-1}}-1\right)\tau(k,l)\cdot\tau(k,l)+\tau(k+1,l)\cdot\tau(k-1,l)=0,\\
&\left(D_{x_1}^4-4D_{x_1}D_{x_3}+3D_{x_2}^2\right)\tau(k,l)\cdot\tau(k,l)=0,
\end{split}
\end{equation}
they have been proved to have the following Gram determinant solutions in Ref.\cite{Han-2017-JPSJ},
\begin{equation}\label{zxe0804}
\tau(k,l)=|m_{ij}(k,l)|_{1\leq i,j\leq N},
\end{equation}
and the elements are given as
\begin{equation*}
\begin{split}
&m_{ij}(k,l)=c_{ij}+\int\phi_i(k,l)\psi_j(k,l)dx_1,
\\&\phi_i(k,l)=(p_i-a)^k \mbox{exp}(\xi_i),
\\&\psi_j(k,l)=(\frac{-1}{q_j+a})^k\mbox{exp}(\tilde{\xi}_j),
\end{split}
\end{equation*}
where $\xi$ and $\tilde{\xi}$ are functions of the variables $x_{-1}, x_1, x_2, x_3$, which can be written as
\begin{equation*}
\begin{split}
&\xi_i=\frac{1}{p_i-a}x_{-1}+p_ix_1+p_i^2x_2+p_i^3x_3+\xi_{i0},\\
&\tilde{\xi}_j=\frac{1}{q_j+a}x_{-1}+q_jx_1-q_j^2x_2+q_j^3x_3+\tilde{\xi}_{i0}.
\end{split}
\end{equation*}

Due to these bilinear equations contain more than two variables, we hope to look for an algebraic solutions satisfying the reduction
\begin{equation}\label{zxe0805}
(\partial_{x_3}-\frac{1}{8}\partial_{x_{-1}}+\frac{1}{4}\partial_{x_1})\tau(k,l)=c\tau(k,l),
\end{equation}
so as to convert the bilinear KP hierarchy into the (1+1)-dimensional bilinear equations
\begin{equation}\label{zxe0806}
\begin{split}
&\left(D^{2}_{x_1}+2aD_{x_1}-D_{x_2}\right)\tau(k+1,l)\cdot\tau(k,l)=0,
\\&\left(D_{x_1}^2+D_{x_1}^4+3D_{x_2}^2-1\right)\tau(k,l)\cdot\tau(k,l)+\tau(k+1,l)\cdot\tau(k-1,l)=0.
\end{split}
\end{equation}

Secondly, by defining $f=\tau(0,0), g=\tau(1,0), h=\tau(-1,0), a=\mbox{i}\alpha$ and with the variables transformation:
\begin{equation*}
x_1=x, x_2=\mbox{i}t
\end{equation*}
Eq.(\ref{zxe0806}) will be transferred into
\begin{equation}\label{zxe0807}
\begin{split}
&\left(D^{2}_{x}+2\mbox{i}\alpha D_{x}-\mbox{i}D_{t}\right)g\cdot f=0,
\\&\left(D_{x}^2+D_{x}^4-3D_{t}^2-1\right)f\cdot f+g\cdot h=0.
\end{split}
\end{equation}
Then, we can obtain the bilinear form of (1+1)-dimensional NLS-Boussinesq equation under the real and complex conjugate condition:
\begin{equation*}
f=\tau(0,0),~~ g=\tau(1,0), ~~g^*=h=\tau(-1,0)
\end{equation*}
Finally, the algebraic solution of (1+1)-dimensional NLS-Boussinesq equation can be presented according to its bilinear form.
\section{The algebraic solution to the (1+1)-dimensional NLS-Boussinesq equation}
This section, we want to construct the algebraic solution from the classical KP hierarchy, that is, the reduction condition Eq.(\ref{zxe0805}) must be satisfied. The detailed steps will be given in the lemma.

\textbf{Lemma} Assume the element of the matrix $m$ is the the following form
\begin{equation*}
m_{kl}^{(\mu\nu n)}=\left(A_k^{(\mu)} B_l^{(\nu)} m^{(n)}\right)\big{|}_{p=\theta, q=\theta^*}
\end{equation*}
where
\begin{equation*}
m^{(n)}=\frac{1}{p+q}\left(-\frac{p-a}{q+a}\right)^ne^{\eta+\tilde{\eta}}, \eta=px_1+p^2x_2, \tilde{\eta}=qx_1-q^2x_2,
\end{equation*}
$\theta$ is a solution of the quadratic dispersion equation
\begin{equation}
3\left(\theta-a\right)^3+6a\left(\theta-a\right)^2+\left(3a^2+\frac{1}{4}\right)(\theta-a)+\frac{1}{8(\theta-a)}=0,
\end{equation}
and $A_k^{(\mu)}, B_l^{(\nu)}$ are two differential operator with respect to $p$ and $q$, defined by
\begin{equation}
\begin{split}
&A_k^{(\mu)}=\sum_{j=0}^ka_j^{(\mu)}\frac{\left[(p-a)\partial_p\right]^{k-j}}{(k-j)!},~~~k\geq0,\\
&B_l^{(\nu)}=\sum_{j=0}^lb_j^{(\nu)}\frac{\left[(q+a)\partial_q\right]^{l-j}}{(l-j)!},~~~l\geq0,
\end{split}
\end{equation}
where the coefficients $a_j^\mu, b_j^\nu$ satisfied
\begin{equation}
\begin{split}
&a_j^{(\mu{+}1)}{=}\sum_{r=0}^{j}\frac{3^{r{+}2}(p{-}a)^3{+}3a2^{r{+}2}(p{-}a)^2{+}(\frac{12a^2{+}1}{4})(p{-}a)+(-1)^{r{+}1}\frac{1}{8(p{-}a)}}{(r{+}2)!}a_{j{-}r}^{(\mu)}, \mu{=}0,1,2\cdots\\
&b_j^{(\nu{+}1)}{=}\sum_{r=0}^{j}\frac{3^{r{+}2}(q{+}a)^3{-}3a2^{r{+}2}(q{+}a)^2{+}(\frac{12a^2{+}1}{4})(q{+}a)+(-1)^{r{+}1}\frac{1}{8(p{-}a)}}{(r{+}2)!}b_{j{-}r}^{(\mu)}, \nu{=}0,1,2\cdots
\end{split}
\end{equation}
then the solution of bilinear equation(\ref{zxe0807}) can be written as
\begin{equation}\label{zxe0811}
\begin{split}
&\tau_n=\underset{1\leq i,j\leq N}{\mbox{det}}=\left|\begin{array}{cccc}
m_{11}^{(N-1,N-1,n)}& m_{13}^{(N-1,N-2,n)}&
\cdots&m_{1,2N-1}^{(N-1,0,n)}\\
m_{31}^{(N-2,N-1,n)}& m_{33}^{(N-2,N-2,n)}&\cdots&m_{3,2N-1}^{(N-2,0,n)}\\
\vdots&\vdots&\vdots&\\
m_{2N-1,1}^{(0,N-1,n)}& m_{2N-1,3}^{(0,N-2,n)}&\cdots&m_{2N-1,2N-1}^{(0,0,n)}
\end{array}\right|
\end{split}
\end{equation}

\textbf{$Proof$}: Based on the solution of the bilinear KP hierarchy (\ref{zxe0804}), let us bring in the functions with the following form
\begin{equation}
\hat{m}^{(n)}=\frac{1}{p+q}(-\frac{p-a}{q+a})^ne^{\gamma+\hat{\gamma}},~~ \hat{\phi}^{(n)}=(p-a)^ne^{\gamma},~~ \hat{\psi}^{(n)}=(\frac{-1}{q+a})^ne^{\hat{\gamma}},
\end{equation}
where
\begin{equation*}
\gamma=\frac{1}{p-a}x_{-1}+px_1+p^2x_2+p^3x_3,~~ \hat{\gamma}=\frac{1}{q+a}+qx_1-q^2x_2+q^3x_3,
\end{equation*}
in addition, these functions should be satisfied the differential form
\begin{equation*}
\begin{split}
&\partial_{x_1}\hat{m}^{(n)}=\hat{\phi}^{(n)}\hat{\psi}^{(n)},\\
&\partial_{x_2}\hat{m}^{(n)}=(\partial_{x_1}\hat{\phi}^{(n)})\hat{\psi}^{(n)}-\hat{\phi}^{(n)}(\partial_{x_1}\hat{\psi}^{(n)}),\\
&\partial_{x_3}\hat{m}^{(n)}=(\partial_{x_1}^2\hat{\phi}^{(n)})\hat{\psi}^{(n)}-(\partial_{x_1}\hat{\phi}^{(n)})\hat{\psi}^{(n)}+\hat{\phi}^{(n)}(\partial_{x_1}^2\hat{\psi}^{(n)}),\\
&\partial_{x_{-1}}\hat{m}^{(n)}=-\hat{\phi}^{(n-1)}\hat{\psi}^{(n+1)},\\
&\hat{m}^{(n+1)}=\hat{m}^{(n)}+\hat{\phi}^{(n)}\hat{\psi}^{(n+1)},\\
&\partial_{x_2}\hat{\phi}^{(n)}=\partial_{x_1}^2\hat{\phi}^{(n)},\\
&\partial_{x_3}\hat{\phi}^{(n)}=\partial_{x_1}^3\hat{\phi}^{(n)},\\
&\hat{\phi}^{(n+1)}=(\partial_{x_1}-a)\hat{\phi}^{(n)},\\
&\partial_{x_2}\hat{\psi}^{(n)}=-\partial_{x_1}^2\hat{\psi}^{(n)},\\
&\partial_{x_3}\hat{\psi}^{(n)}=\partial_{x_1}^3\hat{\psi}^{(n)},\\
&\hat{\psi}^{(n-1)}=-(\partial_{x_1}+a)\hat{\psi}^{(n)}.
\end{split}
\end{equation*}

Then introduce a new entries of the matrix composed by two differential operator:
\begin{equation*}
\hat{m}_{ij}^{(\mu\nu n)}=A_{i}^{(\mu)}B_{j}^{(\nu)}\hat{m}^{(n)}, \hat{\phi}_i^{(\mu n)}=A_{i}^{(\mu)}\hat{\phi}^{(n)},
\hat{\psi}_j^{(\nu n)}=B_{j}^{(\nu)}\hat{\psi}^{(n)}.
\end{equation*}
It is clear that the operators $A_i^{\mu}, B_{j}^{\nu}$ can commute with the differential operator $\partial_{x_1}, \partial_{x_{-1}}, \partial_{x_2}, \partial_{x_3}$, so these functions are suit for the bilinear KP hierarchy (\ref{zxe0803}). Furthermore, for an arbitrary $(i_1, i_2, \cdots i_N;$ $\mu_1, \mu_2, \cdots, \mu_N, j_1, j_2, \cdots, j_N, \nu_1, \nu_2, \cdots, \nu_N)$, the corresponding determinant
\begin{equation*}
\hat{\tau_n}=\mbox{det}\left(\hat{m}_{i_k,j_l}^{(\mu_k, \nu_l,n)}\right)
\end{equation*}
is satisfied the bilinear KP hierarchy, especially, when $\hat{\tau}_n=\underset{1\leq i,j\leq N}{\mbox{det}}\left(\hat{m}_{2i-1,2j-1}^{N-i,N-j,n}\right)$, it is also the solution. Based on the Leibniz rule, one can get
\begin{equation}
\begin{split}
&\left[(p-a)\partial_p\right]^m\left(p^3+\frac{p}{4}-\frac{1}{8(p-a)}\right)\\
=&\sum_{l=0}^{m}\left(
\begin{array}{cc}
m\\l
\end{array}
\right)\left[3^l(p{-}a)^3{+}3a2^l(p{-}a)^2{+}(3a^2{+}\frac{1}{4})(p{-}a){+}({-}1)^{l{+}1}\frac{1}{8(p{-}a)}\right]\left[(p{-}a)\partial_p\right]^{m{-}l}\\
&+\left(a^3+\frac{1}{4}a\right)[(p-a)\partial_p]^m,
\end{split}
\end{equation}
and
\begin{equation}
\begin{split}
&\left[(q+a)\partial_q\right]^m\left(q^3+\frac{q}{4}-\frac{1}{8(q+a)}\right)\\
=&\sum_{l=0}^{m}\left(
\begin{array}{cc}
m\\l
\end{array}
\right)\left[3^l(q{+}a)^3{-}3a2^l(q{+}a)^2{+}(3a^2{+}\frac{1}{4})(q{+}a){+}({-}1)^{l{+}1}\frac{1}{8(q{+}a)}\right]\left[(q{+}a)\partial_q\right]^{m{-}l}\\
&-\left(a^3+\frac{1}{4}a\right)[(q+a)\partial_q]^m.
\end{split}
\end{equation}
Hence, one can obtain the commutator operation
\begin{equation}
\begin{split}
\hspace{-2cm}&\left[A_k^{(\mu)}, p^3+\frac{p}{4}-\frac{1}{8(p-a)}\right]\\
=&\sum_{j=0}^{k}\frac{a_j^{(\mu)}}{(k-j)!}
\left[\left((p-a)\partial_p\right)^{k-j}, p^3+\frac{p}{4}-\frac{1}{8(p-a)} \right]\\
=&\sum_{j=0}^{k-2}\sum_{l=1}^{k{-}j}\frac{a_j^{(\mu)}\left[3^l(p{-}a)^3{+}3a2^l(p{-}a)^2{+}(\frac{12a^2{+}1}{4})(p{-}a){+}\frac{({-}1)^{l{+}1}}{8(p{-}a)}\right]\left[(p{-}a)\partial_p\right]^{k{-}j{-}l}}{l!(k{-}j{-}l)!}\Big|_{p=\theta}\\
\end{split}
\end{equation}
where $\left[,\right]$ devotes the commutator given by $\left[X,Y\right]=XY-YX$.

Suppose $\theta$ is the solution of the quadratic dispersion equation
\begin{equation}
3\left(\theta-a\right)^3+6a\left(\theta-a\right)^2+\left(3a^2+\frac{1}{4}\right)(\theta-a)+\frac{1}{8(\theta-a)}=0,
\end{equation} then the commutator operation equals to zero when $k=0, 1$:
\begin{equation}
\left[A_k^{(\mu)}, p^3+\frac{p}{4}-\frac{1}{8(p-a)}\right]\Bigg|_{p=\theta}=0.
\end{equation}
When $k\geq2$:
\begin{equation}
\begin{split}
\hspace{-2cm}&\left[A_k^{(\mu)}, p^3+\frac{p}{4}-\frac{1}{8(p-a)}\right]\\
=&\sum_{j=0}^{k-2}\sum_{l=2}^{k{-}j}\frac{a_j^{(\mu)}\left[3^l(p{-}a)^3{+}3a2^l(p{-}a)^2{+}(\frac{12a^2{+}1}{4})(p{-}a){+}\frac{({-}1)^{l{+}1}}{8(p{-}a)}\right]\left[(p{-}a)\partial_p\right]^{k{-}j{-}l}}{l!(k{-}j{-}l)!}\Big|_{p=\theta}\\
=&\sum_{j=0}^{k-2}\sum_{\tilde{l}=0}^{k{-}j{-}2}\frac{a_j^\mu\left[3^{\tilde{l}+2}(p{-}a)^3{+}3a2^{\tilde{l}+2}(p{-}a)^2{+}(\frac{12a^2{+}1}{4})(p{-}a){+}\frac{({-}1)^{\tilde{l}{+}1}}{8(p{-}a)}\right]\left[(p{-}a)\partial_p\right]^{k{-}j{-}\tilde{l}{-}2}}{(\tilde{l}{+2})!(k{-}j{-}\tilde{l}{-}2)!}\Big|_{p=\theta}\\
=&\sum_{\hat{j}=0}^{k-2}\left(\sum_{\hat{l}=0}^{\hat{j}}\frac{3^{\hat{l}+2}(p{-}a)^3{+}3a2^{\hat{l}+2}(p{-}a)^2{+}(\frac{12a^2{+}1}{4})(p{-}a){+}\frac{({-}1)^{\hat{l}{+}1}}{8(p{-}a)}}{(\hat{l}+2)!}a_{\hat{j}-\hat{l}}^{(\mu)}\right)\frac{\left((p-a)\partial_p\right)^{k-2-\hat{j}}}{(k-2-\hat{j})!}\Big|_{p=\theta}\\
=&\sum_{\hat{j}=0}^{k-2}a_{\hat{j}}^{(\mu+1)}\frac{\left((p-a)\partial_p\right)^{k-2-\hat{j}}}{(k-2-\hat{j})!}\Big|_{p=\theta}\\
=&A_{k-2}^{(\mu+1)}\big|_{p=\theta}.
\end{split}
\end{equation}
Thus there exists a recurrence relation between the two differential operator
\begin{equation*}
\left[A_{k}^{(\mu)}, p^3+\frac{p}{4}-\frac{1}{8(p-a)}\right]\big|_{p=\theta}=A_{k-2}^{(\mu+1)}\bigg|_{p=\theta},
\end{equation*}
spontaneously, when $k<0$, this operator $A_{k}^{(\mu)}=0$.

Similarly, it is obviously that the differential operator $B_{l}^{(\nu)}$ also satisfies
\begin{equation*}
\left[B_{l}^{\nu},q^3+\frac{q}{4}-\frac{1}{8(q+a)}\right]=B_{l-2}^{(\nu+1)}\big|_{q=\theta^*}
\end{equation*}
when $l>0$, and when $l<0$, we define $B_{l}^{(\nu)}=0.$

Under the above two recurrence equation, the following derivative relation can be derived as:
\begin{equation}
\begin{split}
&\left(\partial_{x_3}+\frac{1}{4}\partial_{x_1}-\frac{1}{8}\partial_{x_{-1}}\right)\hat{m}_{kl}^{(\mu\nu n)}\Big|_{p=\theta, q=\theta^*}\\
=&\left(A_{k}^{(\mu)}B_{l}^{(\nu)}\left(p^3+q^3+\frac{1}{4}(p+q)-\frac{1}{8}(\frac{1}{p-a}+\frac{1}{q+a})\right)\hat{m}^{(n)}\right)\Big|_{p=\theta,q=\theta^*}\\
=&\left(A_{k}^{(\mu)}\left(p^3{+}\frac{p}{4}{-}\frac{1}{8(p{-}a)}\right)B_{l}^{(\nu)}\hat{m}^{(n)}\right)\Big|_{p=\theta,q=\theta^*}
{+}\left(A_{k}^{(\mu)}B_{l}^{(\nu)}\left(q^3{+}\frac{q}{4}{-}\frac{1}{8(q{+}a)}\right)\hat{m}^{(n)}\right)\Big|_{p=\theta,q=\theta^*}\\
=&\left(\left(\left((p^3{+}\frac{p}{4}{-}\frac{1}{8(p{-}a)})A_{k}^{(\mu)}\right){+}A_{k-2}^{(\mu+1)}\right)B_{l}^{(\nu)}\hat{m}^{(n)}\right)\Big|_{p=\theta,q=\theta^*}
\\&{+}\left(A_{k}^{(\mu)}\left(\left(\left(q^3{+}\frac{q}{4}{-}\frac{1}{8(q{+}a)}\right)B_{l}^{(\nu)}\right)\hat{m}^{(n)}\right){+}B_{l-2}^{\nu+1}\right)\Big|_{p=\theta,q=\theta^*}\\
=&\left(p^3{+}\frac{p}{4}{-}\frac{1}{8(p{-}a)}\right)\hat{m}_{kl}^{(\mu\nu n)}\Big|_{p=\theta, q=\theta^*}+\hat{m}_{k-2,l}^{(\mu+1,\nu,n)}\Big|_{p=\theta,q=\theta^*}
\\&+\left(q^3{+}\frac{q}{4}{-}\frac{1}{8(q{+}a)}\right)\hat{m}^{(\mu\nu n)}_{kl}\Big|_{p=\theta,q=\theta^*}+\hat{m}_{k,l-2}^{(\mu,\nu+1,n)}\Big|_{p=\theta,q=\theta^*}.
\end{split}
\end{equation}
Once more, based on the above relation, the differential form of a special determinant rewritten as
\begin{equation}
\hat{\hat{\tau}}_n=\underset{1\leq i,j\leq N}{\mbox{det}}\left(\hat{m}_{2i-1,2j-1}^{N-i,N-j,n}\big|_{p=\theta,q=\theta^*}\right)
\end{equation}
can be work out as
\begin{equation}\label{zxe0821}
\begin{split}
&\left(\partial_{x_3}+\frac{1}{4}\partial_{x_1}-\frac{1}{8}\partial_{x_{-1}}\right)\hat{\hat{\tau}}_n\\
=&\sum_{i=1}^{N}\sum_{j=1}^{N}\triangle_{ij}\left(\partial_{x_3}+\frac{1}{4}\partial_{x_1}-\frac{1}{8}\partial_{x_{-1}}\right)\left(\hat{m}_{2i-1,2j-1}^{N-1,N-j,n}\Bigg|_{p=\theta,q=\theta^*}\right)\\
=&\sum_{i=1}^{N}\sum_{j=1}^{N}\triangle_{ij}\bigg[\left(p^3{+}\frac{p}{4}{-}\frac{1}{8(p{-}a)}\right)\hat{m}_{2i-1,2j-1}^{(N-i,N-j n)}\bigg|_{p=\theta, q=\theta^*}+\hat{m}_{2i-3,2j-1}^{(N-i+1,N-j,n)}\bigg|_{p=\theta,q=\theta^*}
\\&+\left(q^3{+}\frac{q}{4}{-}\frac{1}{8(q{+}a)}\right)\hat{m}^{(N-i,N-j n)}_{2i-1,2j-1}\Bigg|_{p=\theta,q=\theta^*}+\hat{m}_{2i-1,2j-3}^{(N-i,N-j+1,n)}\bigg|_{p=\theta,q=\theta^*}\bigg]\\
=&\left(p^3{+}\frac{p}{4}{-}\frac{1}{8(p{-}a)}\right)N\hat{\hat{\tau}}_n\bigg|_{p=\theta,q=\theta^*}+\sum_{i=1}^{N}\sum_{j=1}^{N}\triangle_{ij}\hat{m}_{2i-3,2j-1}^{(N-i+1,N-j,n)}\bigg|_{p=\theta,q=\theta^*}\\
&+\left(q^3{+}\frac{q}{4}{-}\frac{1}{8(q{+}a)}\right)N\hat{\hat{\tau}}_n\bigg|_{p=\theta,q=\theta^*}+\sum_{i=1}^{N}\sum_{j=1}^{N}\triangle_{ij}\hat{m}_{2i-1,2j-3}^{(N-i,N-j+1,n)}\bigg|_{p=\theta,q=\theta^*},
\end{split}
\end{equation}
where $\triangle_{ij}$ is the $(i,j)$-cofactor of the matrix $\left(\hat{m}_{2i-1,2j-1}^{N-i,N-j,n}\right)$. It is obvious that\\ $\sum_{i=1}^{N}\sum_{j=1}^{N}\triangle_{ij}\hat{m}_{2i-3,2j-1}^{(N-i+1,N-j,n)}\big|_{p=\theta,q=\theta^*}=0$ for $\triangle_{ij}$ is the $(i,j)$-cofactor of the matrix $\left(\hat{m}_{2i-1,2j-1}^{N-i,N-j,n}\right)$ but not the $\left(\hat{m}_{2i-3,2j-1}^{N-i+1,N-j,n}\right)$. Similarly, $\sum_{i=1}^{N}\sum_{j=1}^{N}\triangle_{ij}\hat{m}_{2i-1,2j-3}^{(N{-}i,N{-}j{+}1,n)}\big|_{p=\theta,q=\theta^*}{=}0.$
Therefore, Eq.(\ref{zxe0821}) will be changed into
\begin{equation}\label{zxe0822}
\begin{split}
&\left(\partial_{x_3}+\frac{1}{4}\partial_{x_1}-\frac{1}{8}\partial_{x_{-1}}\right)\hat{\hat{\tau}}_n\\
=&\left(p^3{+}\frac{p}{4}{-}\frac{1}{8(p{-}a)}+q^3{+}\frac{q}{4}{-}\frac{1}{8(q{+}a)}\right)N\hat{\hat{\tau}}_n.
\end{split}
\end{equation}
Due to $\hat{\hat{\tau}}_n$ is a special case of $\hat{\tau}_n$, so $\hat{\hat{\tau}}_n$ is the solution to the (1+1)-dimensional bilinear equation:
\begin{equation}
\begin{split}
&\left(D^{2}_{x_1}+2aD_{x_1}-D_{x_2}\right)\hat{\hat{\tau}}_{n+1}\cdot\hat{\hat{\tau}}_{n}=0,
\\&\left(D_{x_1}^2+D_{x_1}^4+3D_{x_2}^2-1\right)\hat{\hat{\tau}}_{n}\cdot\hat{\hat{\tau}}_{n}+\hat{\hat{\tau}}_{n+1}\cdot\hat{\hat{\tau}}_{n-1}=0.
\end{split}
\end{equation}
Under the reduction Eq.(\ref{zxe0822}), these variables $x_{-1}, x_{3}$ in $\hat{\hat{\tau}}_n$ will become dummy. Thus the matrix entries $\hat{m}_{2N-i,2N-j}^{(N-i,N-j,n)}$ reduce to $m^{(N-i,N-j,n)}_{2N-i,2N-j}$, $\tau_n$ in Eq.(\ref{zxe0811}) satisfy Eq.(\ref{zxe0806}) and the proof is complete.
\section{Complex conjugate condition and regularity}
It has been proved that $\tau_n$ is the solution to the bilinear equation in Lemma, if the variables $x_1, x_2$ satisfy the following reduction
\begin{equation}
x_1=x, x_2=-\mbox{i}t,
\end{equation}
and the variable values of $a_k^{(0)}$ and $b_k^{(0)}$ are complex conjugate each other, then the recurrence values $a_k^{(\mu)}$ and $b_k^{(\mu)}$ will also be complex conjugate each other under the condition $p=\theta, q=\theta^*$, that is
\begin{equation}
b_k^{\mu}|_{q=\theta^*}=(a_k^{(\mu)}|_{p=\theta})^*
\end{equation}
for $\mu=1, 2, \cdots n,$ then the matrix term
\begin{equation}
m_{kj}^{(\mu,\nu,n)*}=m_{kj}^{(\mu,\nu,n)}\big|_{a_k^{(\mu)}\leftrightarrow b_{k}^{(\mu)}, x_2\leftrightarrow-x_2, a\leftrightarrow -a, \theta\leftrightarrow\theta^*}=m_{jk}^{(\mu,\nu, -n)},
\end{equation}
which indicates
\begin{equation}
\tau_n^*=\tau_{-n}.
\end{equation}
It is well known that $f=\tau_0, g=\tau_{1}, g^*=\tau_{-1}$ are the solution of (1+1)-dimensional NLS-Boussinesq equation
\begin{equation}
\begin{split}
&\left(\mbox{i} D_t-2\mbox{i}\alpha D_x-D_x^2\right)g\cdot f=0,\\
&\left(D_x^2+D_x^4-3D_t^2-1\right)f\cdot f+gg^*=0.
\end{split}
\end{equation}
Next, we want to show the rational function $\frac{g}{f}$ is nonsingular. According to the definition of $f=\tau_0$, it is found that $f$ is a determinant of a Hermitian matrix $\tau_0=\mbox{det}(m_{2i-1,2j-1}^{(N-i,N-j,0)})$. In Ref\cite{Ohta-2012-PRS}, it has been proved that when the real part of $p$ is positive, then $\tau_0>0$, conversely, when the real part is negative, $\tau_0<0$. Hence, whether the real part of $p$ is positive or negative, the corresponding function$\tau_0$ is always nonsingular.

Based on the above proof, we can obtain the general high-order rogue wave to the (1+1)-dimensional NLS-Boussinesq equation, which is shown in the Theorem.

\textbf{Theorem} The solution of the (1+1)-dimensional NLS-Boussinesq equation is
\begin{equation}
\begin{split}
&\Phi=e^{\rm i(\alpha x+\alpha^2t)}\frac{\tau_1}{\tau_0},
\\&u=2\frac{\partial^2}{\partial x^2}\mbox{log} \tau_0,
\end{split}
\end{equation}
where
\begin{equation}\label{zxe0830}
\begin{split}
&\tau_n{=}\underset{1\leq i,j\leq N}{\mbox{det}}(m_{2i-1, 2j-1}^{(N-i,N-j,n)}){=}\left|\begin{array}{cccc}
m_{11}^{(N-1,N-1,n)}& m_{13}^{(N-1,N-2,n)}&
\cdots&m_{1,2N-1}^{(N-1,0,n)}\\
m_{31}^{(N-2,N-1,n)}& m_{33}^{(N-2,N-2,n)}&\cdots&m_{3,2N-1}^{(N-2,0,n)}\\
\vdots&\vdots&\vdots&\\
m_{2N-1,1}^{(0,N-1,n)}& m_{2N-1,3}^{(0,N-2,n)}&\cdots&m_{2N-1,2N-1}^{(0,0,n)}
\end{array}\right|,
\end{split}
\end{equation}
with$f=\tau_0, g=\tau_1, g^*=\tau_{-1}$, and the entries of matrix $m_{ij}^{(\mu, \nu, n)}$ are defined by
\begin{equation}
m_{ij}^{(\mu,\nu,n)}=\sum_{k=0}^i\sum_{l=0}^j\frac{a_k^{\mu}}{(i{-}k)!}\frac{a_l^{\nu*}}{(j{-}l)!}\left[(p{-}a)\partial_p\right]^{i{-}k}\left[(q{+}a)\partial_q\right]^{j{-}l}\frac{1}{p{+}q}\left({-}\frac{p{-}a}{q{+}a}\right)e^{(p{+}q)x{-}(p^2{-}q^2)\mbox{i}t}\Bigg|_{p=\theta, q=\theta^*}
\end{equation}
where $\theta$ is the solution of the quadratic dispersion equation
\begin{equation}
3\left(\theta-a\right)^3+6a\left(\theta-a\right)^2+\left(3a^2+\frac{1}{4}\right)(\theta-a)+\frac{1}{8(\theta-a)}=0,
\end{equation}
and $a_{k}^{\mu}$ satisfies the recurrence relation
\begin{equation}
a_j^{(\mu{+}1)}{=}\sum_{r=0}^{j}\frac{3^{r{+}2}(p{-}a)^3{+}3a2^{r{+}2}(p{-}a)^2{+}(\frac{12a^2{+}1}{4})(p{-}a)+(-1)^{r{+}1}\frac{1}{8(p{-}a)}}{(r{+}2)!}a_{j{-}r}^{(\mu)}, \mu{=}0,1,2\cdots
\end{equation}
\section{General high-order rogue wave to NLS-Boussinesq equation}
During the generation of rogue wave, a critical parameter $a$ plays an important role to the pattern of rogue wave. Under the reduction, this parameter is a pure imaginary $\mbox{i}\alpha$. This section, we will discuss the dynamics properties detailly. We first study the solution of the quadratic dispersion equation
\begin{equation}\label{zxe0834}
3\left(p-a\right)^3+6a\left(p-a\right)^2+\left(3a^2+\frac{1}{4}\right)(p-a)+\frac{1}{8(p-a)}=0,
\end{equation}
this equation includes four roots as follows when $a=i\alpha$:
\begin{equation}
\begin{split}
&p_1{=}\frac{\mbox{i}\alpha}{2}{+}\frac{k_4^{\frac{1}{2}}}{12}
{+}\frac{1}{12}\left(\frac{36\mbox{i}\alpha}{k_4^{\frac{1}{2}}}{-}\frac{144\alpha^4{-}24\alpha^2{+}73}{k_2^{\frac{1}{3}}}{-}24\alpha^2{-}k_2^{\frac{1}{3}}{-}4\right)^{\frac{1}{2}},\\
&p_2{=}\frac{\mbox{i}\alpha}{2}{+}\frac{k_4^{\frac{1}{2}}}{12}
{-}\frac{1}{12}\left(\frac{36\mbox{i}\alpha}{k_4^{\frac{1}{2}}}{-}\frac{144\alpha^4{-}24\alpha^2{+}73}{k_2^{\frac{1}{3}}}{-}24\alpha^2{-}k_2^{\frac{1}{3}}{-}4\right)^{\frac{1}{2}},\\
&p_3{=}\frac{\mbox{i}\alpha}{2}{-}\frac{k_4^{\frac{1}{2}}}{12}
{+}\frac{1}{12}\left(\frac{36\mbox{i}\alpha}{{-}k_4^{\frac{1}{2}}}{-}\frac{144\alpha^4{-}24\alpha^2{+}73}{k_2^{\frac{1}{3}}}{-}24\alpha^2{-}k_2^{\frac{1}{3}}{-}4\right)^{\frac{1}{2}},\\
&p_4{=}\frac{\mbox{i}\alpha}{2}{-}\frac{k_4^{\frac{1}{2}}}{12}
{-}\frac{1}{12}\left(\frac{36\mbox{i}\alpha}{{-}k_4^{\frac{1}{2}}}{-}\frac{144\alpha^4{-}24\alpha^2{+}73}{k_2^{\frac{1}{3}}}{-}24\alpha^2{-}k_2^{\frac{1}{3}}{-}4\right)^{\frac{1}{2}},\\
\end{split}
\end{equation}
where
\begin{equation}
\begin{split}
k&=3456\alpha^6-2592\alpha^4+2952\alpha^2-1058,\\ k_1&=-1728\alpha^6+432\alpha^4-1332\alpha^2-215,\\
k_2&=k_1+18k^{\frac{1}{2}},\\
k_3&=144\alpha^4-12\alpha^2k_2^{\frac{1}{3}}+k_2^{\frac{2}{3}}-24\alpha^2-2k_2^{\frac{1}{3}}+73,\\
k_4&=\frac{k_3}{k_2^{\frac{1}{3}}}.
\end{split}
\end{equation}
Due to the complexity of the solutions, we cannot divide the real part from the corresponding solutions. So we give the first-order rogue wave based on the Eq.(\ref{zxe0830}), and discuss the effect of the parameter $\alpha$ on the pattern of the rogue wave. For simplicity, let $a_0^{(0)}=1, a_{1}^{(0)}=0, N=1$, then the functions $f$ and $g$ can be written as
\begin{equation}\label{zxe0837}
\begin{split}
f&=\frac{e^{2\kappa(2\gamma t+x)}\left[\left(\kappa x+2\gamma\kappa t-\frac{1}{2}\right)^2+(2\kappa^2t)^2+\frac{1}{4}\right]\left[(\alpha-\gamma)^2+\kappa^2\right]}{2\kappa^3},\\
g&=\frac{\left[\left(m_1\kappa x{+}2\kappa m_3t{+}\frac{\gamma{-}\alpha{-}\kappa{+}2i\kappa}{2}\right)^2{+}\left(m_2\kappa x{+}2\kappa m_4t{+}\frac{\gamma{-}\alpha{+}\kappa{+}2i\kappa}{2}\right)^2{+}\frac{\kappa^2{+}\left(\alpha{+}\gamma\right)^2}{2}\right]\left(i\kappa{-}\gamma{+}\alpha\right)}{4\kappa^3\left(i\kappa{+}\gamma{-}\alpha\right)}
\end{split}
\end{equation}
\begin{equation*}
\begin{split}
m_1&=\alpha+\kappa-\gamma,\\
m_2&=\alpha-\kappa-\gamma,\\
m_3&=\kappa^2+2\kappa\gamma-\gamma^2+\alpha\gamma-\alpha\kappa,\\
m_4&=\kappa^2-2\kappa\gamma-\gamma^2+\alpha\gamma+\alpha\kappa,
\end{split}
\end{equation*}
where $\kappa$ is the real part of $p$ and $\gamma$ is the imaginary part. Then the first-order rogue wave solution in NLS-Boussinesq equation is
\begin{equation}
\begin{split}
\Phi&=e^{\rm i(\alpha x+\alpha^2t)}\frac{g}{f},
\\u&=2\frac{\partial^2}{\partial x^2}\mbox{log} f
\end{split}
\end{equation}
where $f, g$ is in Eq.(\ref{zxe0837}).

With the simplify calculation, the modular square of the short-wave component $|\Phi|^2$ has five critical points:
\begin{equation}
\begin{split}
(x_1, t_1)&=(\frac{1}{2\kappa}, 0),\\
(x_2, t_2)&=\left(\frac{\left(\alpha-2\gamma\right)\left(3\varpi^2-\kappa^2\right)^{\frac{1}{2}}+\varpi^2+\kappa^2}{2\kappa\left(\varpi^2+\kappa^2\right)},\frac{\left(3\varpi^2-\kappa^2\right)^{\frac{1}{2}}}{4\kappa\left(\varpi^2+\kappa^2\right)}\right),\\
(x_3, t_3)&=\left(\frac{\left(2\gamma-\alpha\right)\left(3\varpi^2-\kappa^2\right)^{\frac{1}{2}}+\varpi^2+\kappa^2}{2\theta\left(\varpi^2+\kappa^2\right)},\frac{\left(3\varpi^2-\kappa^2\right)^{\frac{1}{2}}}{-4\kappa\left(\varpi^2+\kappa^2\right)}\right),\\
(x_4, t_4)&=\left(\frac{\left(\alpha\gamma-\gamma^2+\kappa^2\right)\left(3\kappa^2-\varpi^2\right)^{\frac{1}{2}}+\kappa\left(\varpi^2+\kappa^2\right)}{2\kappa^2\left(\varpi^2+\kappa^2\right)},\frac{\left(\gamma-\alpha\right)\left(3\kappa^2-\varpi^2\right)^{\frac{1}{2}}}{4\kappa^2\left(\varpi^2+\kappa^2\right)}\right),\\
(x_5, t_5)&=\left(\frac{\left(\gamma^2-\alpha\gamma-\kappa^2\right)\left(3\kappa^2-\varpi^2\right)^{\frac{1}{2}}+\kappa\left(\varpi^2+\kappa^2\right)}{2\kappa^2\left(\varpi^2+\kappa^2\right)},-\frac{\left(\gamma-\alpha\right)\left(3\kappa^2-\varpi^2\right)^{\frac{1}{2}}}{4\kappa^2\left(\varpi^2+\kappa^2\right)}\right).
\end{split}
\end{equation}
where $\varpi=\alpha-\gamma$. With the Hessian matrix with two variables at these critical points, the first-order cofactor is
\begin{equation}
\begin{split}
H_1(x, t)&=\left[\frac{\partial^2|\Phi|^2}{\partial x^2}\right],\\
H_1(x_1, t_1)&=\frac{192\kappa^4\left[(\alpha-\gamma)^2-\kappa^2\right]}{\left[(\alpha-\gamma)^2+\kappa^2\right]}\\
H(x_2, t_2)=H(x_3, t_3)&=-\frac{6\kappa^4\left[(\alpha-\gamma)^2+\kappa^2\right]}{\left(\alpha-\gamma\right)^4}\\
H(x_4, t_4)=H(x_5, t_5)&=6\left[(\alpha-\gamma)^2+\kappa^2\right]
\end{split}
\end{equation}
and the second-order cofactor is
\begin{equation}
\begin{split}
H(x, t)&=\left[\frac{\partial^2|\Phi|^2}{\partial x^2}\frac{\partial^2|\Phi|^2}{\partial t^2}-\left(\frac{\partial^2|\Phi|^2}{\partial x\partial t}\right)^2\right]\\
H(x_1, t_1)&=\frac{16384\kappa^{10}\left[(\alpha-\gamma)^2-3\kappa^2\right]\left[3(\alpha-\gamma)^2-\kappa^2\right]}{\left[(\alpha-\gamma)^2+\kappa^2\right]^4},\\
H(x_2, t_2)=H(x_3, t_3)&=\frac{64\kappa^{10}\left[3(\alpha-\gamma)^2-\kappa^2\right]\left[(\alpha-\gamma)^2+\kappa^2\right]^2}{\left(\gamma-\alpha\right)^{10}},\\
H(x_4, t_4)=H(x_5, t_5)&=-64\left[(\alpha-\gamma)^2+\kappa^2\right]^2\left[(\alpha-\gamma)^2-3\kappa^2\right].
\end{split}
\end{equation}
Again, we begin to discuss the solution of Eq.(\ref{zxe0834}). we only consider the positive value of parameter $\alpha$ because of the negative value has the similar property.

When $k{>}0$, that is $\alpha{>}\left(\frac{\left(107{+}51\sqrt{17}\right)^{\frac{1}{3}}}{12}{-}\frac{8}{3\left(107{+}51\sqrt{17}\right)^{\frac{1}{3}}}{+}\frac{1}{4}\right)^{\frac{1}{2}}$, the imaginary part of $p_1$ is $\frac{\alpha}{2}{+}\frac{\left(-k_4\right)^{\frac{1}{2}}}{12}$the real part is $\frac{1}{12}\left(\frac{36\mbox{i}\alpha}{\sqrt{k_4}}-\frac{144\alpha^4-24\alpha^2+73}{k_2^{\frac{1}{3}}}-24\alpha^2-k_2^{\frac{1}{3}}-4\right)^{\frac{1}{2}}$ , and $p_2$ is the conjugate number of $p_1$, but we cannot give the exact real part and imaginary part for both $p_3$ and $p_4$.

When $k{<}0$, that is $0\leq\alpha{<}\left(\frac{\left(107{+}51\sqrt{17}\right)^{\frac{1}{3}}}{12}{-}\frac{8}{3\left(107{+}51\sqrt{17}\right)^{\frac{1}{3}}}{+}\frac{1}{4}\right)^{\frac{1}{2}}$
these four roots $p_1, p_2, p_3, p_4$ will not be separated as real part and imaginary part explicitly. We only can study the effect of parameter $\alpha$ from the graph. Based on the analysis of the critical values, we get a conclusion that, if $p=p_2$ or $p=p_4$, there will appear three patterns of rogue waves: four-petals state, dark state, bright state. if $p=p_1$ or $p=p_3$, there only exist two patterns: four-petals state and bright state. So the following rogue wave is presented on the choice of $p=p_2$. The approximate classification is:

(a) Four-petals state($0\leq\alpha<0.1796$), in this case, $(x_2, t_2), (x_3, t_3)$ are the two local maximums, $(x_4, t_4), (x_5, t_5)$ are the two local minimums, $(x_1, t_1)$ is not a local extremum.

(b) Dark state($0.1796\leq\alpha<0.6538$), in this case, $(x_1, t_1)$ is the only local minimum and $(x_2, t_2), (x_3, t_3)$ are two local maximums.

(c) Bright state($\alpha\geq0.6538$), in this case, $(x_1, t_1)$ is the local maximum, $(x_4, t_4), (x_5, t_5)$ are two local minimums.

\begin{figure}[H]
\subfigure[]{\includegraphics[height=0.2\textwidth]{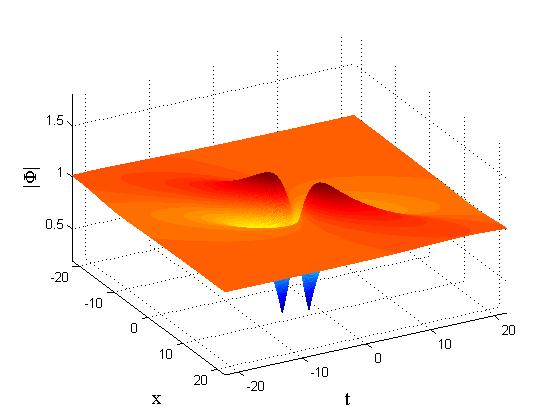}}
\centering
\subfigure[]{\includegraphics[height=0.2\textwidth]{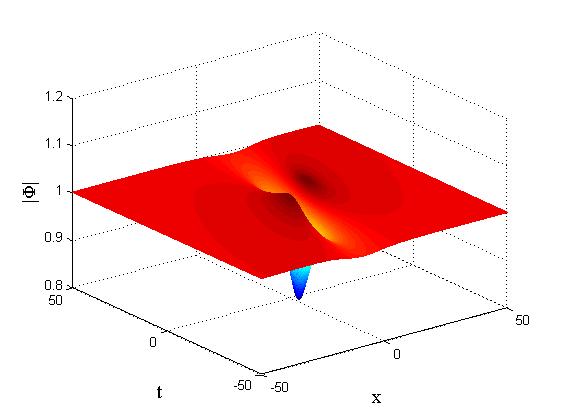}}
\centering
\subfigure[]{\includegraphics[height=0.2\textwidth]{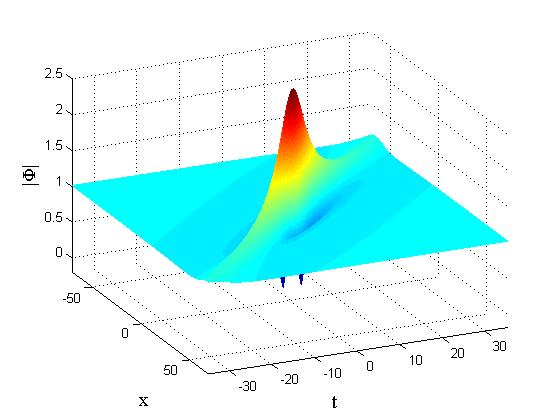}}
\centering
\caption{\small(Color online) The first-order rogue wave of NLS-Boussinesq equation: (a) Four-petals state $\alpha=0$, (b) Dark state $\alpha=\frac{1}{2}$, (c) Bright state $\alpha=\frac{\sqrt{3}}{2}$.}
\label{zxe08fig1}
\end{figure}
\begin{figure}[H]
\subfigure[]{\includegraphics[height=0.2\textwidth]{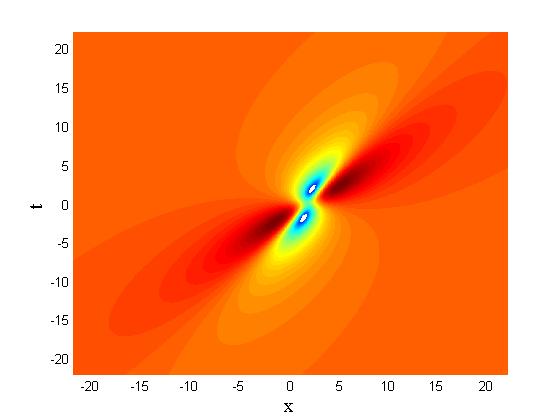}}
\centering
\subfigure[]{\includegraphics[height=0.2\textwidth]{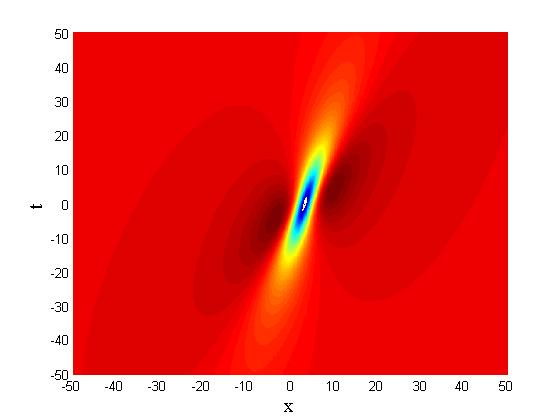}}
\centering
\subfigure[]{\includegraphics[height=0.2\textwidth]{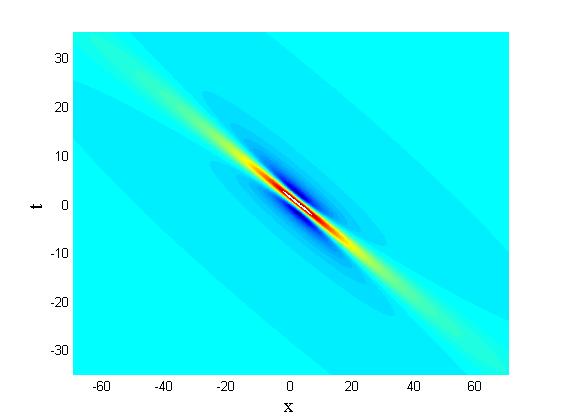}}
\centering
\caption{\small(Color online) The corresponding density plots of Fig. (\ref{zxe08fig1}).}
\label{zxe08fig2}
\end{figure}
\begin{figure}[H]
\subfigure[]{\includegraphics[height=0.2\textwidth]{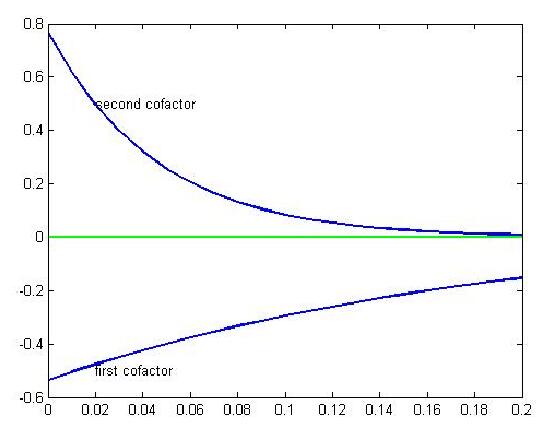}}
\centering
\subfigure[]{\includegraphics[height=0.2\textwidth]{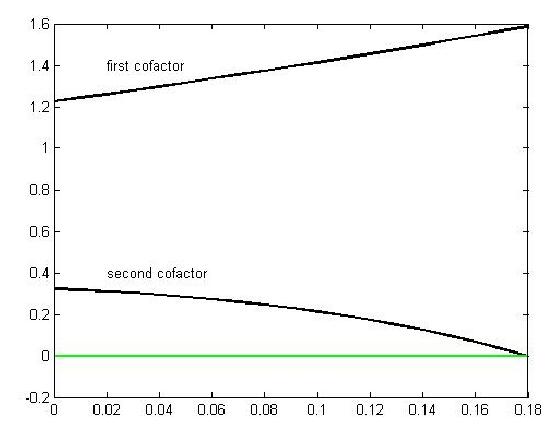}}
\centering
\subfigure[]{\includegraphics[height=0.2\textwidth]{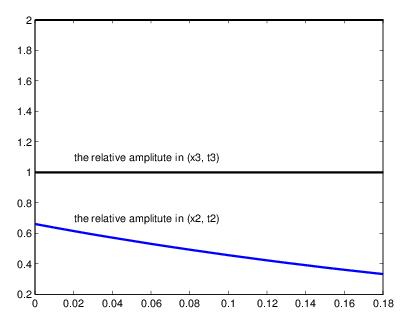}}
\centering
\caption{\small(Color online) The reason for the generation of four-petals state : (a) the cofactor at $(x_2, t_2), (x_3, t_3)$ points altered by parameter $\alpha$, (b) the cofactor at $(x_4, t_4), (x_5, t_5)$ points altered by parameter $\alpha$, (c) the relative amplitude at these four critical points.}
\label{zxe08fig3}
\end{figure}
\begin{figure}[H]
\subfigure[]{\includegraphics[height=0.2\textwidth]{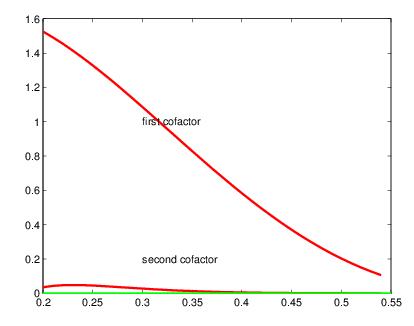}}
\centering
\subfigure[]{\includegraphics[height=0.2\textwidth]{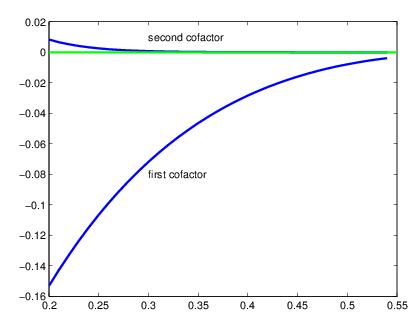}}
\centering
\subfigure[]{\includegraphics[height=0.2\textwidth]{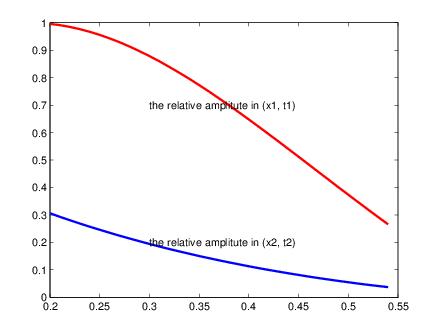}}
\centering
\caption{\small(Color online) The reason for the generation of dark state: (a) the cofactor at $(x_1, t_1)$ point altered by parameter $\alpha$, (b) the cofactor at $(x_2, t_2), (x_3, t_3)$ points altered by parameter $\alpha$, (c) the relative amplitude at these three critical points.}
\label{zxe08fig4}
\end{figure}
\begin{figure}[H]
\subfigure[]{\includegraphics[height=0.2\textwidth]{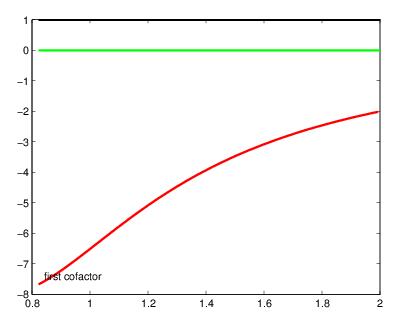}}
\centering
\subfigure[]{\includegraphics[height=0.2\textwidth]{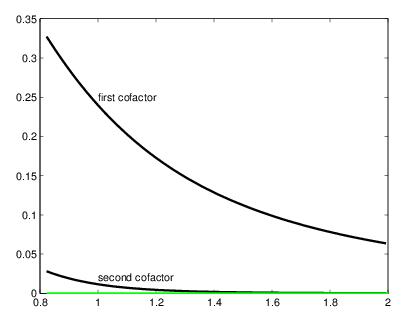}}
\centering
\subfigure[]{\includegraphics[height=0.2\textwidth]{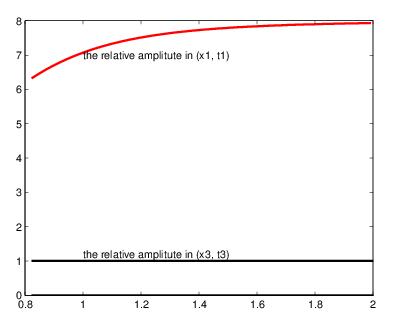}}
\centering
\caption{\small(Color online) The reason for the generation of bright state: (a) the cofactor at $(x_1, t_1)$ point altered by parameter $\alpha$, (b) the cofactor at $(x_4, t_4), (x_5, t_5)$ points altered by parameter $\alpha$, (c) the relative amplitude at these three critical points.}
\label{zxe08fig5}
\end{figure}

Furthermore, as to these three kinds of patterns, the amplitudes at these critical points are exhibited in Fig.\ref{zxe08fig3}, \ref{zxe08fig4}, \ref{zxe08fig5} respectively.  We mainly discuss the relative amplitude because of the rogue wave is irritated by the $1-plane$ background wave.
In Fig.\ref{zxe08fig3}, (a) indicates that these two critical points $(x_2, t_2), (x_3, t_3)$ are the maximums, (b) shows that these two critical points $(x_4, t_4), (x_5, t_5)$ are minimums. In addition, the relative amplitude at points $(x_2, t_2), (x_3, t_3)$ are larger than the points $(x_4, t_4), (x_5, t_5)$, which is called the four-petals state.

Then the generation mechanism of dark rogue wave is depicted in Fig.\ref{zxe08fig4}
It is clear that, the critical point $(x_1, t_1)$ is the minimum, $(x_2, t_2), (x_3, t_3)$ are the maximums, and the relative amplitude at $(x_1, t_1)$ is larger than at $(x_2, t_2), (x_3, t_3)$, which generates the dark rogue wave.
The analysis to the bright state is the same as the dark state, due to the order of magnitude to the second cofactor in $(x_1, t_1)$is too large to exhibit. (a)indicates the critical point $(x_1, t_1)$ is the maximum, $(x_4, t_4), (x_5, t_5)$ are the minimums, and the relative amplitude at $(x_1, t_1)$ is shorter than at $(x_4, t_4), (x_5, t_5)$, which generates the bright rogue wave.

It is must be emphasized that, when $\alpha{<}\left(\frac{\left(107{+}51\sqrt{17}\right)^{\frac{1}{3}}}{12}{-}\frac{8}{3\left(107{+}51\sqrt{17}\right)^{\frac{1}{3}}}{+}\frac{1}{4}\right)^{\frac{1}{2}}$, there exist two patterns, and as the parameter $\alpha$ gets bigger, the rogue wave changes from the four-petals state to dark state. Meanwhile, when $\alpha{>}\left(\frac{\left(107{+}51\sqrt{17}\right)^{\frac{1}{3}}}{12}{-}\frac{8}{3\left(107{+}51\sqrt{17}\right)^{\frac{1}{3}}}{+}\frac{1}{4}\right)^{\frac{1}{2}}$, it only appears the bright rogue wave, but the four-petals state cannot exist. Maybe in this case, if there will four critical points, then it can obtain the four-petals rogue wave, evolved from the bright state.
\section{High-order rogue wave}
\begin{figure}[H]
\subfigure[]{\includegraphics[height=0.2\textwidth]{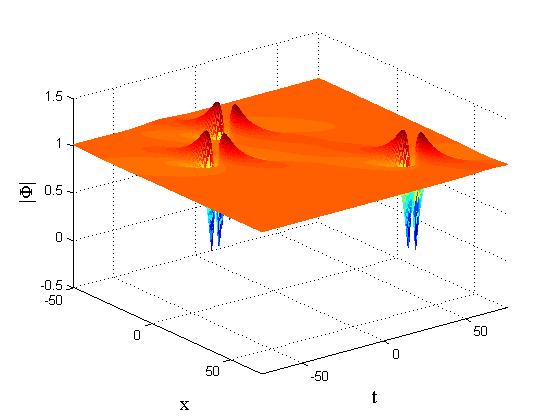}}
\centering
\subfigure[]{\includegraphics[height=0.2\textwidth]{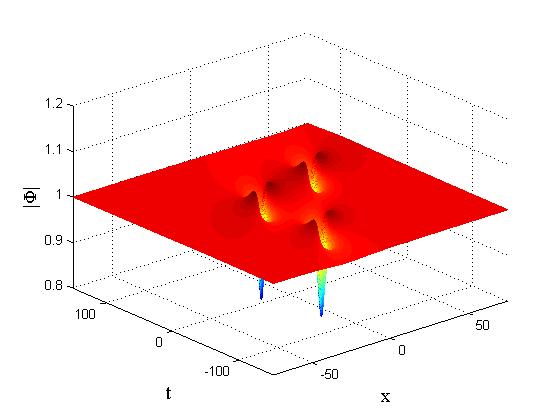}}
\centering
\subfigure[]{\includegraphics[height=0.2\textwidth]{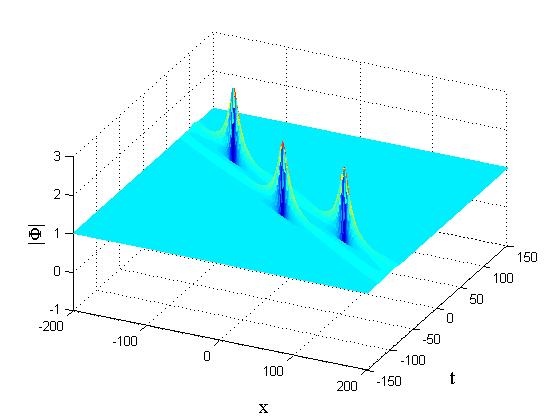}}
\centering
\caption{\small(Color online)The second-order rogue wave of NLS-Boussinesq equation, (a) the four-petals state for $\alpha=0, a_{3}^{(0)}=500$, (b) Dark state for $\alpha=\frac{1}{2}, a_{3}^{(0)}=1000$, (c) the bright state for $\alpha=\frac{\sqrt{3}}{2}, a_{3}^{(0)}=100+10\mbox{i}$.}
\label{zxe08fig6}
\end{figure}

\begin{figure}[H]
\subfigure[]{\includegraphics[height=0.2\textwidth]{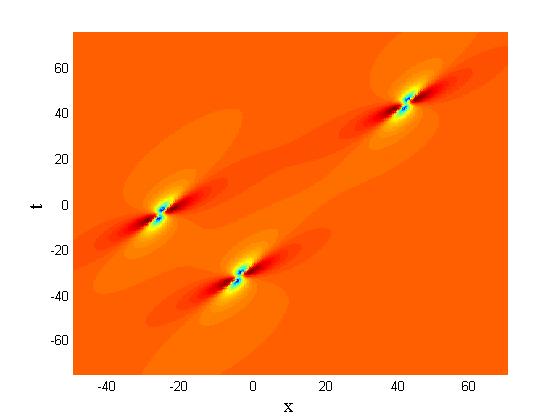}}
\centering
\subfigure[]{\includegraphics[height=0.2\textwidth]{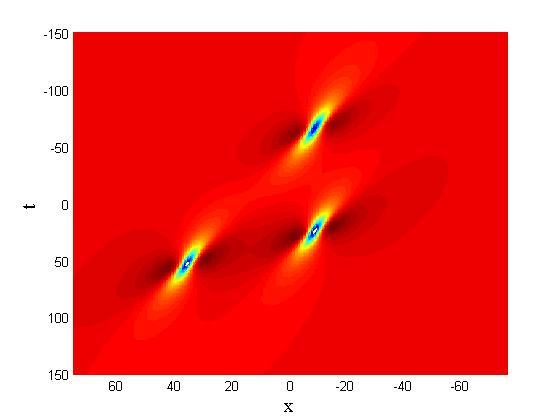}}
\centering
\subfigure[]{\includegraphics[height=0.2\textwidth]{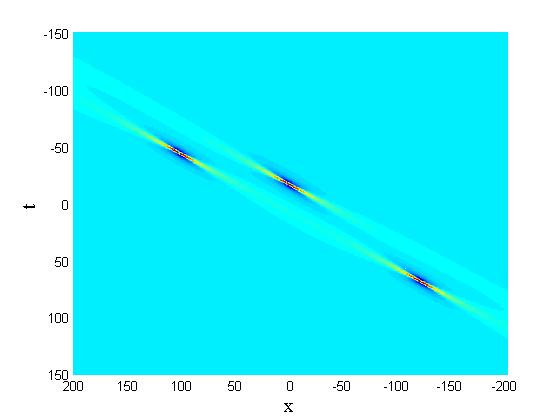}}
\centering
\caption{\small(Color online) The corresponding density plots of Fig.(\ref{zxe08fig6})}
\label{zxe08fig7}
\end{figure}

The second order rogue wave can be obtained from Eq.(\ref{zxe0830}) by taking $N=2$, and the initial values is on the choice of $a_0^{(0)}=1, a_1^{(0)}=a_2^{(0)}=0$, then the function $f, g$ will be written as
\begin{equation}
\begin{split}
f=\left|\begin{array}{cc}
m_{11}^{(1,1,0)}& m_{13}^{(1,0,0)}\\
m_{31}^{(0,1,0)}& m_{33}^{(0,0,0)}
\end{array}\right|,
~~~~~g=\left|\begin{array}{cc}
m_{11}^{(1,1,1)}& m_{13}^{(1,0,1)}\\
m_{31}^{(0,1,1)}& m_{33}^{(0,0,1)}
\end{array}\right|
\end{split}
\end{equation}
where
\begin{equation}
\begin{split}
m_{i,j}^{(\mu,\nu,0)}&=A_{i}^{(\mu)}B_{j}^{(\nu)}\frac{1}{p+q}e^{(p+q)x-(p^2-q^2)\mbox{i}t}\big|_{p=\theta, q=\theta^*},\\
m_{i,j}^{(\mu,\nu,1)}&=-A_{i}^{(\mu)}B_{j}^{(\nu)}\frac{1}{p+q}\frac{p-\mbox{i}\alpha}{q+\mbox{i}\alpha}e^{(p+q)x-(p^2-q^2)\mbox{i}t}\big|_{p=\theta, q=\theta^*}.\\
\end{split}
\end{equation}
Similarly, the second-order rogue wave has also three patterns: four-petals state, dark state, bright state, which are depicted in Fig. \ref{zxe08fig6}
Its corresponding density plots are shown in Fig.\ref{zxe08fig7}

These three patterns second-order rogue waves( four-petals state, dark state, bright state) all contain three fundamental one-order rogue waves and the structures of these three fundamental one-order rogue wave exhibit triangle arrays. In addition, the patterns affected by the parameter $\alpha$ are similar to the fundamental one-order rogue wave, that is, when $0\leq\alpha<0.1796$, the second-order rogue wave is four-petals state, when $0.1796\leq\alpha<0.6538$, it is the dark rogue wave, when $\alpha>0.6538$, it is bright rogue wave.

\begin{figure}[H]
\subfigure[]{\includegraphics[height=0.2\textwidth]{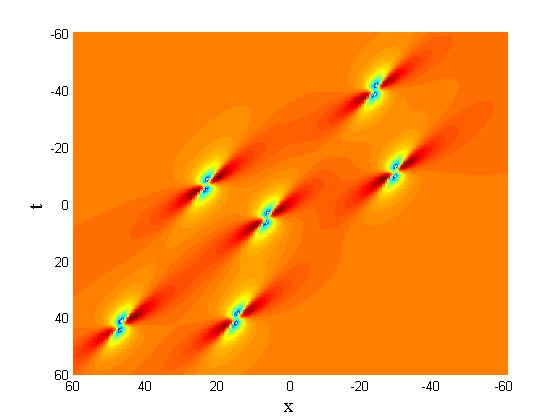}}
\centering
\subfigure[]{\includegraphics[height=0.2\textwidth]{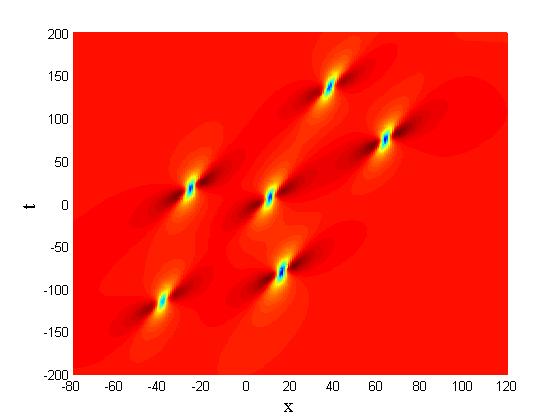}}
\centering
\caption{\small(Color online)The third-order rogue waves: (a) four-petals state with the choices $\alpha=0, a_{0}^{(0)}=1, a_{1}^{(0)}=a_{2}^{(0)}=a_{3}^{(0)}=a_{4}^{(0)}=0, a_{5}^{(0)}=1000$, (b) dark state by choosing $\alpha=\frac{1}{2}, a_{0}^{(0)}=1, a_{1}^{(0)}=a_{2}^{(0)}=a_{3}^{(0)}=a_{4}^{(0)}=0, a_{5}^{(0)}=300000$.}
\label{zxe08fig8}
\end{figure}

Finally, the four-petals state and dark state of third-order rogue wave are presented in Fig.(\ref{zxe08fig8}) by choosing some appropriate initial values. But the bright third-order and higher-order rogue waves cannot be exhibited due to the expressions are too complicated to illustrate here.
\section{Conclusion}
In this paper, We construct the general high-order rogue wave by using KP reduction technique and analyze the dynamical property of rogue waves. The obtained rogue waves exhibit three patterns: four-petals state, dark state and bright state under the extreme value theory, which is governed by a free parameter $\alpha$. We mainly analysis the Hessian matrix of the function $|\Phi|$ with respect to the variables $x$ and $t$ from two cases $k>0$ and $k<0$. When $k>0$, if the function has two maximums and two minimums, it will appear the four-petals state. As the parameter $\alpha$ evolution, the upper relative amplitude is becoming smaller and the lower relative amplitude is bigger, when the critical points number reduces to three, two maximum values and one minimum, there appear the dark state. When $k<0$, the function always has two minimum values and one maximum no matter what the values of parameter $\alpha$, it appears the bright rogue wave. In ref. \cite{Zhao-2014-PRE1}, Zhao et. al. studied the transition between the four-petals state and the bright state or the dark state as the change of the relative frequency. But in our paper, the transformation between the bright state and four-petals state are not realized, maybe it is owing to the upper relative amplitude is smaller than the lower relative amplitude to the four-petals state, which is different from the four-petals state in ref. \cite{Zhao-2014-PRE1}. This analysis can be also used to the dynamical behavior of high-order rogue wave for the reason that the high-order rogue wave is the superposition of fundamental rogue wave, such as, the second-order is consisted of three fundamental rogue wave and the third-order contains six fundamental rogue wave.

As we all know that, by using the Darboux transformation, the interactions between high-order rogue wave and breather or bright-dark soliton are discussed. But as to the Hirota bilinear method and KP reduction technique, there are scarcely any studies about the interaction between the high-order rogue wave and other localized wave. This paper we have obtained the high order rogue wave to the NLS-Boussinesq equation, we should try to focus on the hybrid solutions in the future topic.

\end{document}